\documentclass[10pt]{article}
\usepackage{fullpage}
\usepackage{amsmath}
\usepackage{amssymb}
\usepackage{amsfonts}
\usepackage[dvips]{epsfig}
\usepackage{color}

\def\##1{\underline{#1}}
\def\=#1{\underline{\underline{#1}}}

\def\+#1{\underline{\bf #1}}
\def\*#1{\underline{\underline{\bf #1}}}

\def\r#1{(\ref{#1})}
\def\l#1{\label{#1}}
\def\c#1{\cite{#1}}

\def\le{\left(}
\def\ri{\right)}
\def\les{\left[}
\def\ris{\right]}
\def\lec{\left\{}
\def\ric{\right\}}

\def\.{\mbox{ \tiny{$^\bullet$} }}

\def\eps{\varepsilon}

\begin{document}

\begin{center}

\LARGE{ {\bf Gain and loss enhancement in active and passive particulate composite materials 
}}
\end{center}

\begin{center}
\vspace{10mm} \large

 Tom G. Mackay\footnote{E--mail: T.Mackay@ed.ac.uk.}\\
{\em School of Mathematics and
   Maxwell Institute for Mathematical Sciences\\
University of Edinburgh, Edinburgh EH9 3FD, UK}\\
and\\
 {\em NanoMM~---~Nanoengineered Metamaterials Group\\ Department of Engineering Science and Mechanics\\
Pennsylvania State University, University Park, PA 16802--6812,
USA}\\
 \vspace{3mm}
 Akhlesh  Lakhtakia\footnote{E--mail: akhlesh@psu.edu}\\
 {\em NanoMM~---~Nanoengineered Metamaterials Group\\ Department of Engineering Science and Mechanics\\
Pennsylvania State University, University Park, PA 16802--6812, USA}

\normalsize

\end{center}

\begin{center}
\vspace{15mm} {\bf Abstract}

\end{center}

Two active dielectric  materials may be blended together to realize a  homogenized composite material (HCM) which exhibits more  gain than either component material. Likewise,
two dissipative dielectric  materials may be blended together to realize an HCM which exhibits more loss than either  component material. Sufficient conditions for such gain/loss enhancement  were established using the Bruggeman homogenization formalism. Gain/loss enhancement  arises when  (i) the imaginary parts of the relative permittivities of both component materials are similar in magnitude and (ii) the real parts of the relative permittivities of both component materials are dissimilar in magnitude.

\vspace{14mm}


\vspace{14mm}

\section{Introduction}

Two (or more) particulate materials may be mixed together to realize a homogenized composite material (HCM), provided that the particles making up the component materials are much smaller than the wavelengths involved \c{L96}. To be of practical value, an HCM is generally required to exhibit a desirable blend of certain properties  of its component materials. Metamaterials are HCMs whose performances  exceed those of their component materials \c{Walser,Meta}. Within the electromagnetic realm, many instances of such HCMs can be found.
For examples: through the process of homogenization,  the phenomenon of weak nonlinearity may be enhanced \c{Boyd96,Liao,L01}, and the group speed may be enhanced
beyond the maximum group speed in the component materials
 \c{Solna2,ML_group_vel_1} or weakened below  the minimum group speed in the component materials \c{Gao_06}.

In this short article, the prospect of enhancing gain by means of homogenization is explored for HCMs arising from active component materials. The dual process of loss enhancement in HCMs arising from dissipative component materials is also considered. The  well--established Bruggeman homogenization formalism \c{Br,Aspnes,MAEH} is employed, all component materials being thereby treated on the same footing. Accordingly, this formalism is applicable for all values of volume fraction of the component materials.

\section{Homogenization via the Bruggeman formalism}

Let us consider a composite  material comprising two distinct  materials labelled `a' and  `b'
that are distributed randomly as electrically small spheres.
 Both component materials  are isotropic dielectric materials with relative permittivities $\eps_a = \eps^r_a +i \eps^i_a $ and $\eps_b = \eps^r_b +i \eps^i_b$, respectively, wherein  $\eps^{r,i}_{a,b} \in \mathbb{R} $ and $\eps_a \neq \eps_b$. Physical plausibility requires the imposition
 of the restriction $\eps^r_a \eps^r_b >0 $ on the Bruggeman formalism \c{ML_Br_limitation}.

 The Bruggeman estimate $\eps_{Br} = \eps^r_{Br} +i \eps^i_{Br}$ of the HCM relative permittivity
 is provided implicitly by the quadratic equation \c{Br}
 \begin{equation} \l{Br_eqn}
 2 \eps^2_{Br} + \eps_{Br} \les \eps_a \le 1-3f_a \ri + \eps_b \le 3f_a - 2 \ri \ris - \eps_a \eps_b =0\,,
 \end{equation}
 with $f_a$ being the volume fraction of component material `a'.
The limiting conditions $\eps_{Br} \to \eps_b$ as $f_a \to 0$, and $\eps_{Br} \to \eps_a$ as $f_a \to 1$
allow the correct root to be extracted from Eq.~\r{Br_eqn}.

When both component materials are active (i.e., $\eps^i_{a,b} < 0$),  the phenomenon of gain enhancement is signified by
$ \eps^i_{Br} < \min\left\{\eps^i_{a},\eps^i_{b}\right\}$.  When both component materials are dissipative (i.e., $\eps^i_{a,b} > 0$),  the phenomenon of loss enhancement is signified by
$ \eps^i_{Br} > \max\left\{\eps^i_{a},\eps^i_{b}\right\}$.

\begin{figure}[!htb]
\begin{center}
\begin{tabular}{c}
\includegraphics[width=14.0cm]{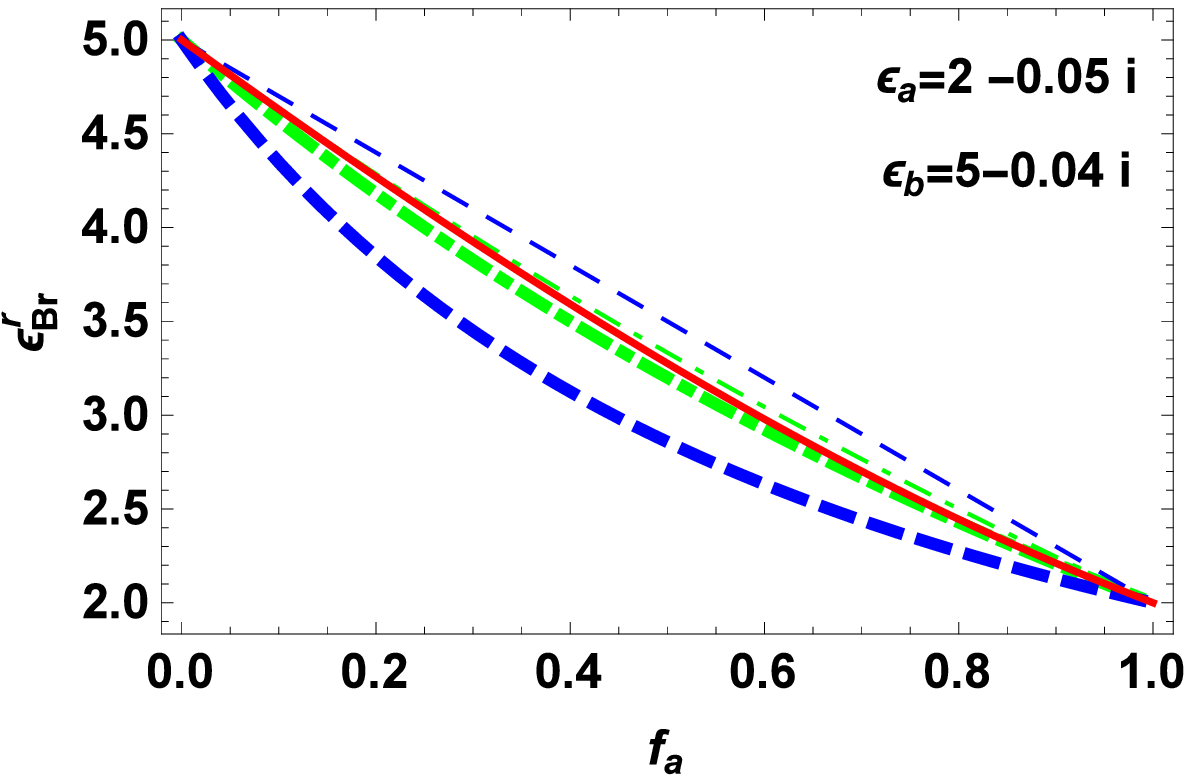}\\
\includegraphics[width=14.0cm]{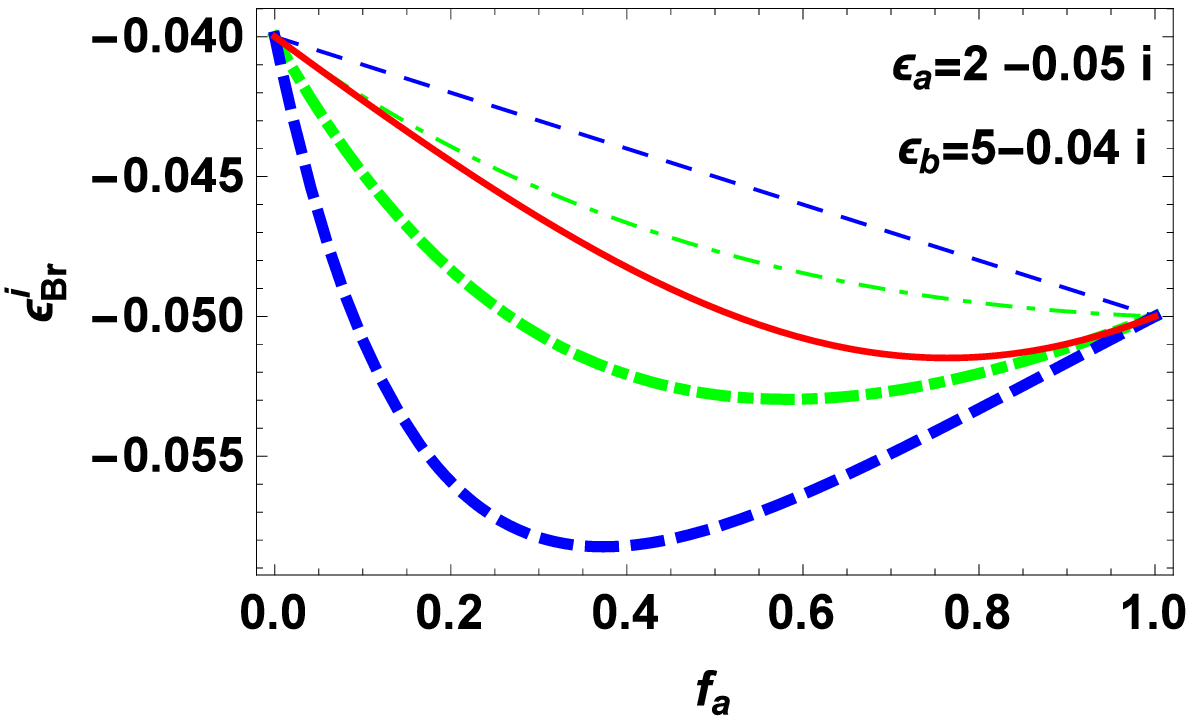}
\end{tabular}
\end{center}
 \caption{The real and imaginary parts of the HCM relative permittivity  $\eps_{Br}$ estimated by the Bruggeman formalism (red, solid curves) plotted against volume fraction $f_a$, when
 $\eps_a = 2- 0.05 i$ and $\eps_a = 5- 0.04 i$. Also plotted are the Hashin--Shtrikman bounds: $\mbox{HS}_\alpha$ (thin, green, broken dashed curves) and $\mbox{HS}_\beta$ (thick, green, broken dashed curves);  and the Wiener bounds:  $\mbox{W}_\alpha$ (thin, blue, dashed curves)
 and $\mbox{W}_\beta$ (thick, blue, dashed curves).
 } \label{fig1}
\end{figure}

To illustrate the phenomenon of gain enhancement, let us consider a specific example. Suppose that the component materials are active ones, specified by $\eps_a = 2 - 0.05 i$ and $\eps_b = 5 - 0.04 i$. The real and imaginary parts   of the Bruggeman estimate of the  HCM relative permittivity are plotted against volume fraction in Fig.~\ref{fig1}.
Also plotted in this figure are two well--established bounds on the HCM relative permittivity, namely the
Wiener bounds
\c{Wiener}
\begin{equation}
\left.
\begin{array}{l}
\mbox{W}_\alpha = f_a \eps_a + f_b \eps_b
\vspace{8pt} \\
\mbox{W}_\beta = \displaystyle{ \le \frac{f_a}{\eps_a} + \frac{f_b}{\eps_b} \ri^{-1}}
\end{array}
\right\}\,\l{Wiener}
\end{equation}
 and the Hashin--Shtrikman
bounds \c{HS}
\begin{equation}
\left.
\begin{array}{l}
\displaystyle{\mbox{HS}_\alpha = \eps_b + \frac{3 f_a \eps_b \le \eps_a - \eps_b
\ri}
{\eps_a + 2 \eps_b - f_a \le \eps_a - \eps_b \ri}}  \vspace{8pt} \\
\displaystyle{\mbox{HS}_\beta = \eps_a + \frac{3 f_b \eps_a \le \eps_b - \eps_a
\ri} {\eps_b + 2 \eps_a - f_b \le \eps_b - \eps_a \ri}}
\end{array}
\right\}\,. \l{Hashin}
\end{equation}
Herein, $f_b = 1- f_a$ is the volume fraction of component material `b'.
Originally, the Wiener bounds and the Hashin--Shtrikman bounds were
 derived for HCMs characterized by wholly real--valued constitutive parameters,
but generalizations to complex--valued constitutive parameters later emerged \c{Milton}.

The Hashin--Shtrikman bound $\mbox{HS}_\alpha$ is
equivalent to the Maxwell Garnett  estimate of the HCM relative
permittivity,  based on the homogenization of a random dispersal of spheres of component
material `a' embedded in the host component material `b', valid for $f_a  \lesssim 0.3$ \c{MG}.  Similarly,
$\mbox{HS}_\beta$ is equivalent to the Maxwell Garnett  estimate
of the HCM relative permittivity, based on the homogenization of a random dispersal of spheres  of component material `b'
 embedded in the host component
material `a', valid  for $f_b \lesssim 0.3$.

The real part of $\eps_{Br}$ is seen in Fig.~\ref{fig1} to decrease uniformly from $\eps^r_b$ to $\eps^r_a$ as $f_a$ increases from $0$ to $1$. Furthermore, $\eps^r_{Br}$ is tightly bounded by $\mbox{HS}_\alpha$ and $\mbox{HS}_\beta$, and   less tightly bounded by $W_\alpha$ and $W_\beta$. The imaginary part of $\eps_{Br}$ follows a more interesting trajectory as $f_a$ increases:
$\eps^i_{Br}$ decreases from $\eps^i_b$ at $f_a =0$, reaches  a minimum value at $f_a \approx 0.8$, and then increases to reach $\eps^i_a$ at $f_a =1$. Thus, according to the Bruggeman formalism, gain enhancement arises in the vicinity of $f_a \approx 0.8$, with the minimum value of $\eps^i_{Br}$  ($\approx -0.0515$) being approximately $3\%$ smaller than $\min\left\{\eps^i_{a},\eps^i_{b}\right\}$.
Furthermore,
${\rm Im}\left(\mbox{HS}_\beta\right) <\min\left\{\eps^i_{a},\eps^i_{b}\right\}$ when $0.7 \lesssim f_a < 1$. Thus, gain enhancement
is also predicted by the Maxwell Garnett formalism.

\begin{figure}[!htb]
\begin{center}
\begin{tabular}{c}
\includegraphics[width=14.0cm]{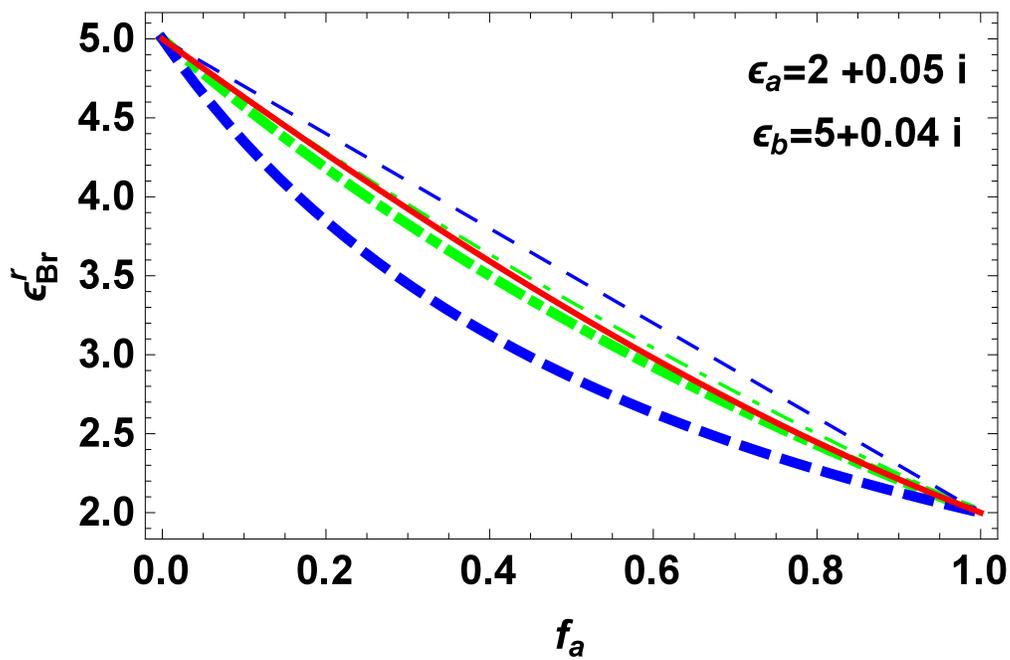}\\
\includegraphics[width=14.0cm]{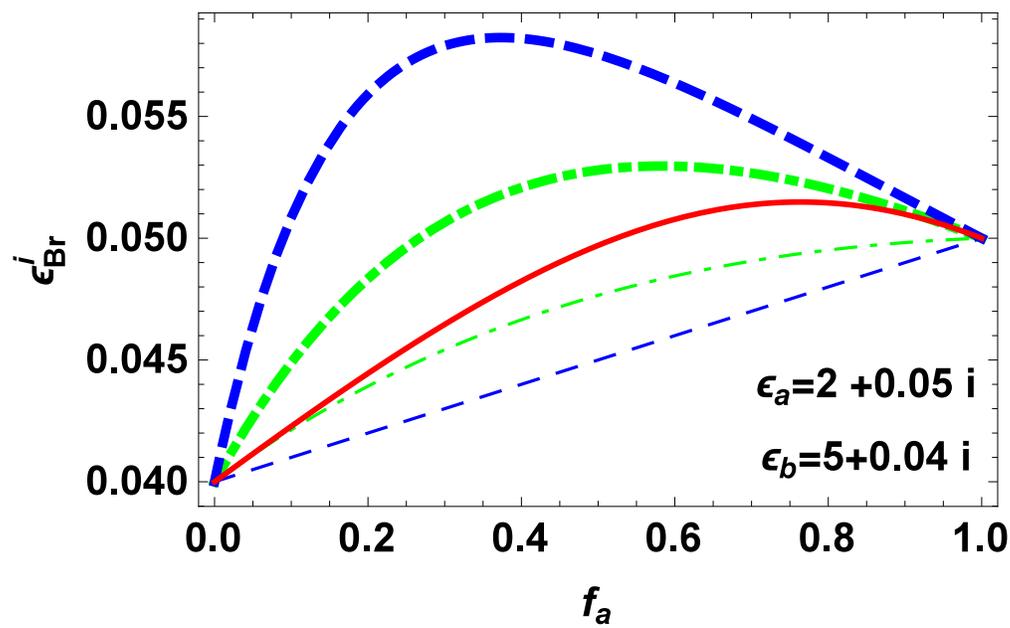}
\end{tabular}
\end{center}
 \caption{As Fig.~\ref{fig1} except that both component materials are dissipative,  having the relative permittivities
 $\eps_a = 2+ 0.05 i$ and $\eps_a = 5+ 0.04 i$.
 } \label{fig2}
\end{figure}

Loss enhancement mirrors  gain enhancement. To support this assertion, let us consider the dissipative counterpart of the active HCM considered in Fig.~\ref{fig1}. In Fig.~\ref{fig2}, plots are presented which are equivalent to those presented
in Fig.~\ref{fig1} but now the component materials are dissipative ones, specified by $\eps_a = 2 + 0.05 i$ and $\eps_b = 5 + 0.04 i$. As in Fig.~\ref{fig1},  $\eps_{Br}^r$ in Fig.~\ref{fig2} decreases uniformly from $\eps^r_b$ to $\eps^r_a$ as $f_a$ increases from $0$ to $1$; moreover, $\eps^r_{Br}$ is tightly bounded by $\mbox{HS}_\alpha$ and $\mbox{HS}_\beta$, and  less tightly bounded by $W_\alpha$ and $W_\beta$. The plot of   $\eps_{Br}^i$ in Fig.~\ref{fig2} displays loss enhancement with
the maximum value  of $\eps^i_{Br}$ ($\approx 0.0515$) being approximately $3\%$ larger than $\max\left\{\eps^i_{a},\eps^i_{b}\right\}$. In addition, ${\rm Im}\left(\mbox{HS}_\beta\right) >\max\left\{\eps^i_{a},\eps^i_{b}\right\}$ when $0.7 \lesssim f_a < 1$. Thus, loss enhancement
is  predicted by both the Bruggeman formalism and the Maxwell Garnett formalism.

\begin{figure}[!htb]
\begin{center}
\begin{tabular}{c}
\includegraphics[width=14.0cm]{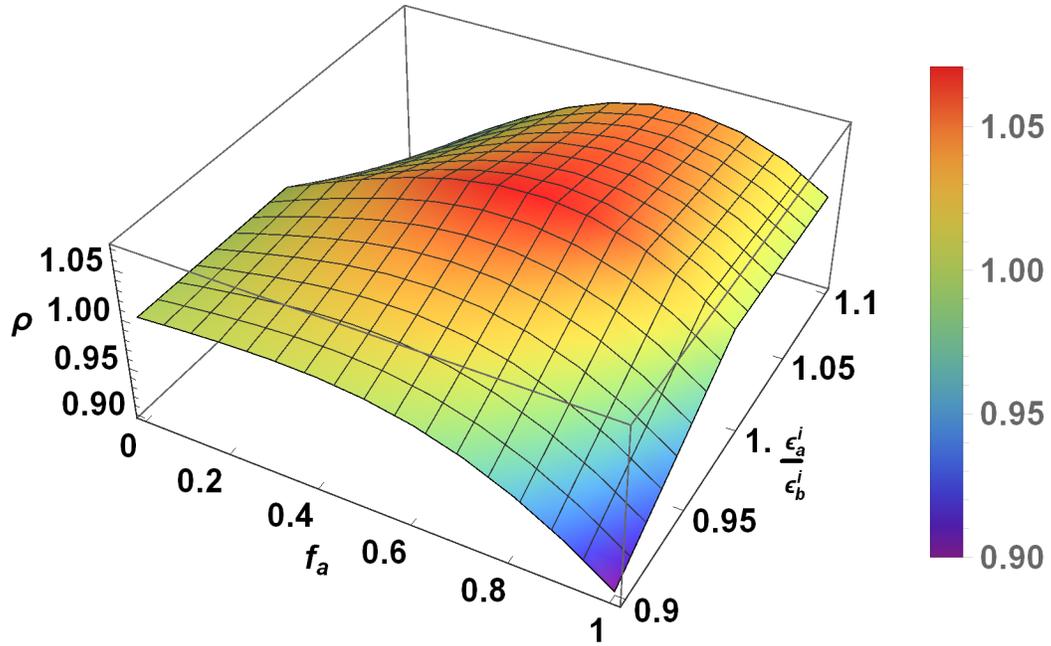}
\end{tabular}
\end{center}
 \caption{The gain--enhancement index $\rho$ plotted against  $f_a\in[0,1]$ and $\eps^i_a / \eps^i_b\in[0.9,1.1]$, when   $\eps^r_a = 2$, $\eps^r_b = 5$, and
 $\eps^i_b = -0.04$.
\label{fig3}}
\end{figure}

Since the active and dissipative scenarios effectively represent two different sides of the same coin, henceforth in this section we focus
 on gain enhancement. Let us now turn to  the gain--enhancement index
\begin{equation}
\rho = \frac{\eps^i_{Br}}{\min \lec \eps^i_a, \eps^i_b \ric }
\end{equation}
estimated using the Bruggeman formalism. Gain enhancement is signified by $\rho > 1$.
For $\eps^r_a = 2$, $\eps^r_b = 5$, and $\eps^i_b = -0.04$,  $\rho$ is plotted against volume fraction $f_a$ and the ratio
$\eps^i_a/\eps^i_b$ in Fig.~\ref{fig3}. Gain enhancement is evident for mid--ranges value of $f_a$ when   $\eps^i_a/\eps^i_b\simeq1$. Specifically for this particular example,
\begin{itemize}
\item[(a)] $\rho$ is as high as about $1.05$, with its maximum value
occurring for $f_a \approx 0.6$ and $
\eps^i_a/\eps^i_b = 1$; and
\item[(b)] there is no gain enhancement  for $
\eps^i_a/\eps^i_b  \lesssim  0.95$ and for  $
\eps^i_a/\eps^i_b  \gtrsim  1.07$, regardless of the value of $f_a$.
\end{itemize}

\begin{figure}[!htb]
\begin{center}
\begin{tabular}{c}
\includegraphics[width=14.0cm]{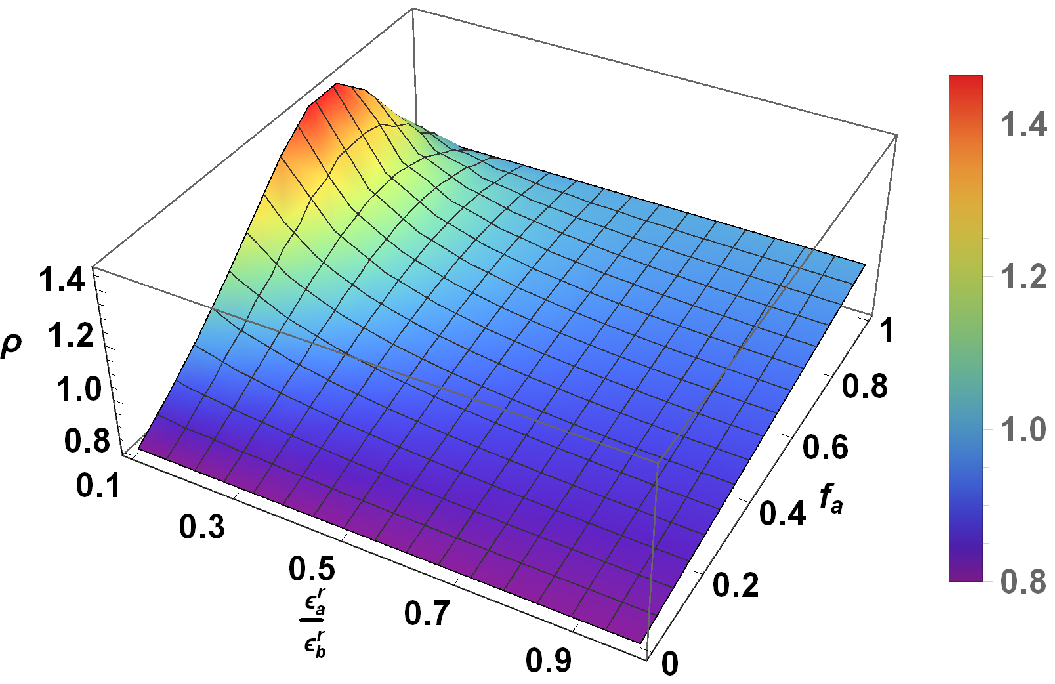}\\
\includegraphics[width=14.0cm]{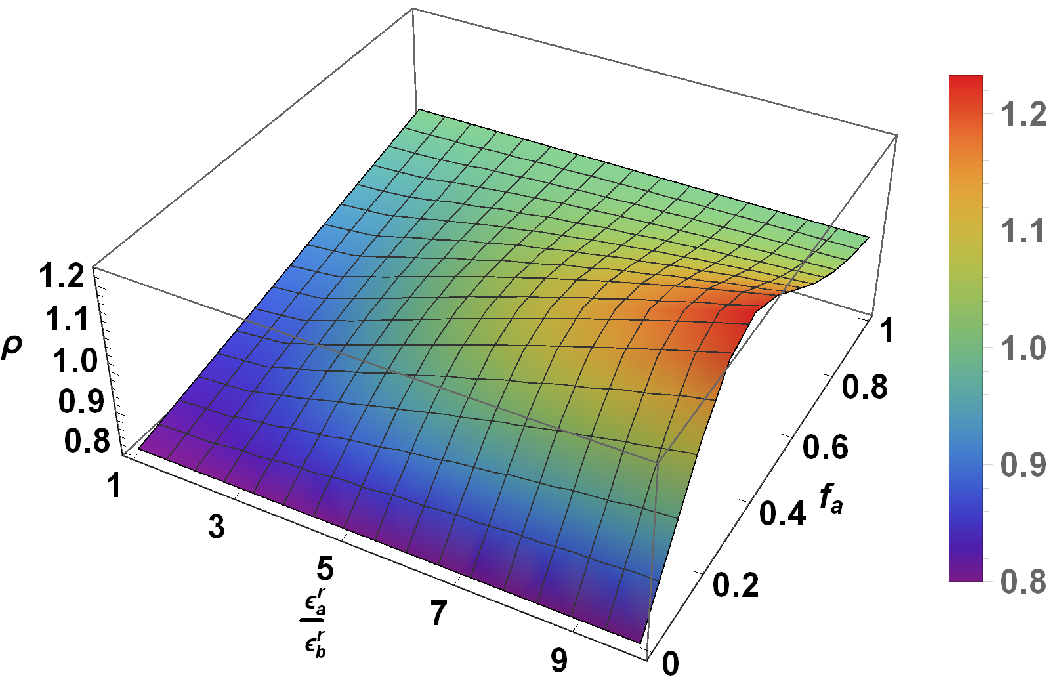}
\end{tabular}
\end{center}
 \caption{The gain--enhancement index $\rho$ plotted against  $f_a\in[0,1]$ and $\eps^r_a / \eps^r_b\in[0.9,1]\cup[1,10]$, when   $\eps^i_a = -0.05$, $\eps^i_b = -0.04$,
and
 $\eps^r_b = 5$.
 } \label{fig4}
\end{figure}

The dependency of $\rho$ upon $\eps_a^r$ and $\eps_b^r$ is delineated in Fig.~\ref{fig4}, wherein $\rho$ is plotted against $f_a$ and $\eps^r_a/\eps^r_b$ for $\eps^i_a = -0.05$, $\eps^i_b = -0.04$, and $\eps^r_b = 5$. As in Fig.~\ref{fig3},  $\rho$ is high for mid-range values of $f_a$ when the ratio $\eps^r_a/\eps^r_b$ deviates most from unity in Fig.~\ref{fig4}. Specifically  for this particular example,
\begin{itemize}
\item[(a)] $\rho$ is as high as about $1.4$, with its maximum value
occurring for $f_a \approx 0.7$ and   $
\eps^r_a/\eps^r_b = 0.1$;
\item[(b)] $\rho$ is as high as about $1.2$, with its maximum value
occurring for $f_a \approx 0.5$ and   $
\eps^r_a/\eps^r_b  = 10$; and
\item[(c)] there is no gain enhancement for
$ \eps^r_a/\eps^r_b\approx  1$, regardless of the value of $f_a$.
\end{itemize}

 \section{Conditions for gain/loss enhancement}

 The foregoing and  similar calculations led us to conclude that gain enhancement
should be expected when
\begin{itemize}
\item[(i)] $\eps^i_a < 0$ and $\eps^i_b<0$,
\item[(ii)] the ratio $\eps^i_a/\eps^i_b$ is close to unity, and
\item[(iii)] the ratio $\eps^r_a/\eps^r_b$ is either very small or very large.
\end{itemize}
Loss enhancement should be expected when $\eps^i_a >0$, $\eps^i_b>0$, and the
conditions (ii) and (iii) are satisfied.
In order to formally establish this understanding soundly, we used
the Bruggeman equation \r{Br_eqn}  to obtain the gradient
 \begin{equation} \l{dBr}
 \frac{d \eps_{Br}}{d f_a} = \frac{3 \eps_{Br} \le \eps_a - \eps_b \ri}{4 \eps_{Br} + \eps_a \le 1-3f_a \ri + \eps_b \le 3f_a - 2 \ri}\,.
 \end{equation}
 This expression underlies further analysis.

\subsection{Gain enhancement} \l{Sec_gain_enhancement}

 Suppose that both component materials are active, i.e.,  $\eps^i_a <0 $ and $ \eps^i_b<0$.
If  $\eps^i_a \geq  \eps^i_b$, then a \textit{sufficient} condition for gain enhancement is that the
 gradient
 \begin{equation} \l{c1}
\lim_{f_a \to 0}   \frac{d \eps^i_{Br} }{d f_a}   < 0\,.
 \end{equation}
Given that
\begin{equation}
\lim_{f_a \to 0}    \eps_{Br} =\eps_b\,,
\end{equation}
Eq.~\r{dBr} yields
 \begin{equation}
\lim_{f_a \to 0}   \frac{d \eps^i_{Br} }{d f_a}  = \frac{3 \eps_b \le \eps_a - \eps_b \ri}{2 \eps_b + \eps_a  }\,,
 \end{equation}
and hence
\begin{equation}
\lim_{f_a \to 0}   \frac{d \eps^i_{Br} }{d f_a}
 = \frac{ 3 \lec \les \eps^i_b \le  \eps^r_a - \eps^r_b \ri + \eps^r_b \le \eps^i_a - \eps^i_b \ri \ris
  \le 2 \eps^r_b + \eps_a^r \ri
  -  \les  \eps^r_b \le  \eps^r_a - \eps^r_b \ri - \eps^i_b\le \eps^i_a - \eps^i_b \ri  \ris \le 2 \eps^i_b + \eps^i_a \ri \ric}{\le 2 \eps^r_b + \eps_a^r \ri^2 + \le 2 \eps^i_b + \eps^i_a \ri^2 }\,.
\end{equation}
The sufficient condition \r{c1} for gain enhancement  is therefore logically equivalent to
\begin{eqnarray} \l{cond1}
&& \les \eps^i_b \le  \eps^r_a - \eps^r_b \ri + \eps^r_b \le \eps^i_a - \eps^i_b \ri \ris
  \le 2 \eps^r_b + \eps_a^r \ri
  <  \les  \eps^r_b \le  \eps^r_a - \eps^r_b \ri - \eps^i_b\le \eps^i_a - \eps^i_b \ri  \ris \le 2 \eps^i_b + \eps^i_a \ri \,.
\end{eqnarray}

If  $\eps^i_b \geq  \eps^i_a$ then a \textit{sufficient} condition for gain enhancement is that the
 gradient
 \begin{equation} \l{c2}
\lim_{f_a \to 1}   \frac{d \eps^i_{Br} }{d f_a}  > 0\,.
 \end{equation}
 Following the same argument as used to derive condition \r{cond1}, we found that the  sufficient condition \r{c2} for gain enhancement
 is logically equivalent to
\begin{eqnarray} \l{xcond1}
&& \les \eps^i_a \le  \eps^r_a - \eps^r_b \ri + \eps^r_a \le \eps^i_a - \eps^i_b \ri \ris
  \le 2 \eps^r_a + \eps_b^r \ri
  >  \les  \eps^r_a \le  \eps^r_a - \eps^r_b \ri - \eps^i_a\le \eps^i_a - \eps^i_b \ri  \ris \le 2 \eps^i_a + \eps^i_b \ri .
\end{eqnarray}

The special case $ \eps^i_a = \eps^i_b$ is noteworthy. Both the conditions \r{cond1} and \r{xcond1} then reduce to
\begin{equation} \l{cond3}
 \le  \eps^r_a - \eps^r_b \ri^2 >0.
\end{equation}
Since condition \r{cond3} is always satisfied because $\eps_a\ne\eps_b$,  gain enhancement is guaranteed for all values of $ \eps^r_a$ and $ \eps^r_b$, provided that $ \eps^i_a = \eps^i_b$.

 \subsection{Loss enhancement}

 Suppose both component materials are dissipative, i.e.,  $\eps^i_a > 0 $ and $ \eps^i_b>0$.
If  $\eps^i_a \leq  \eps^i_b$, then a sufficient condition for loss enhancement is that the gradient
 \begin{equation}
\lim_{f_a \to 0}   \frac{d \eps^i }{d f_a}   > 0\,,
 \end{equation}
which, in the manner described in \S\ref{Sec_gain_enhancement}, is logically equivalent to the condition
\begin{eqnarray} \l{cond4}
&& \les \eps^i_b \le  \eps^r_a - \eps^r_b \ri + \eps^r_b \le \eps^i_a - \eps^i_b \ri \ris
  \le 2 \eps^r_b + \eps_a^r \ri
  >  \les  \eps^r_b \le  \eps^r_a - \eps^r_b \ri - \eps^i_b\le \eps^i_a - \eps^i_b \ri  \ris \le 2 \eps^i_b + \eps^i_a \ri \,.
\end{eqnarray}
If $\eps^i_b \leq  \eps^i_a$, then a sufficient condition for loss enhancement is that the gradient
 \begin{equation}
\lim_{f_a \to 1}   \frac{d \eps^i }{d f_a}    < 0\,,
 \end{equation}
which is logically equivalent to the condition
\begin{eqnarray} \l{cond5}
&& \les \eps^i_a \le  \eps^r_a - \eps^r_b \ri + \eps^r_a \le \eps^i_a - \eps^i_b \ri \ris
  \le 2 \eps^r_a + \eps_b^r \ri
  <  \les  \eps^r_a \le  \eps^r_a - \eps^r_b \ri - \eps^i_a\le \eps^i_a - \eps^i_b \ri  \ris \le 2 \eps^i_a + \eps^i_b \ri \,.
\end{eqnarray}

As in \S\ref{Sec_gain_enhancement},  both conditions \r{cond4} and \r{cond5} reduce to condition \r{cond3} for the special case $ \eps^i_a = \eps^i_b$. Therefore,
 loss enhancement is guaranteed for all values of $ \eps^r_a$ and $ \eps^r_b$ when $ \eps^i_a = \eps^i_b$.

\subsection{Numerical illustration}

The conditions \r{cond1}  and \r{xcond1}  provide a convenient means of exploring the parameter space of the relative permittivities of the component materials that support gain enhancement, and  conditions \r{cond4} and \r{cond5} play the same role for  loss enhancement. Let us illustrate this assertion with a numerical example.

\begin{figure}[!htb]
\begin{center}
\begin{tabular}{c}
\includegraphics[width=12.0cm]{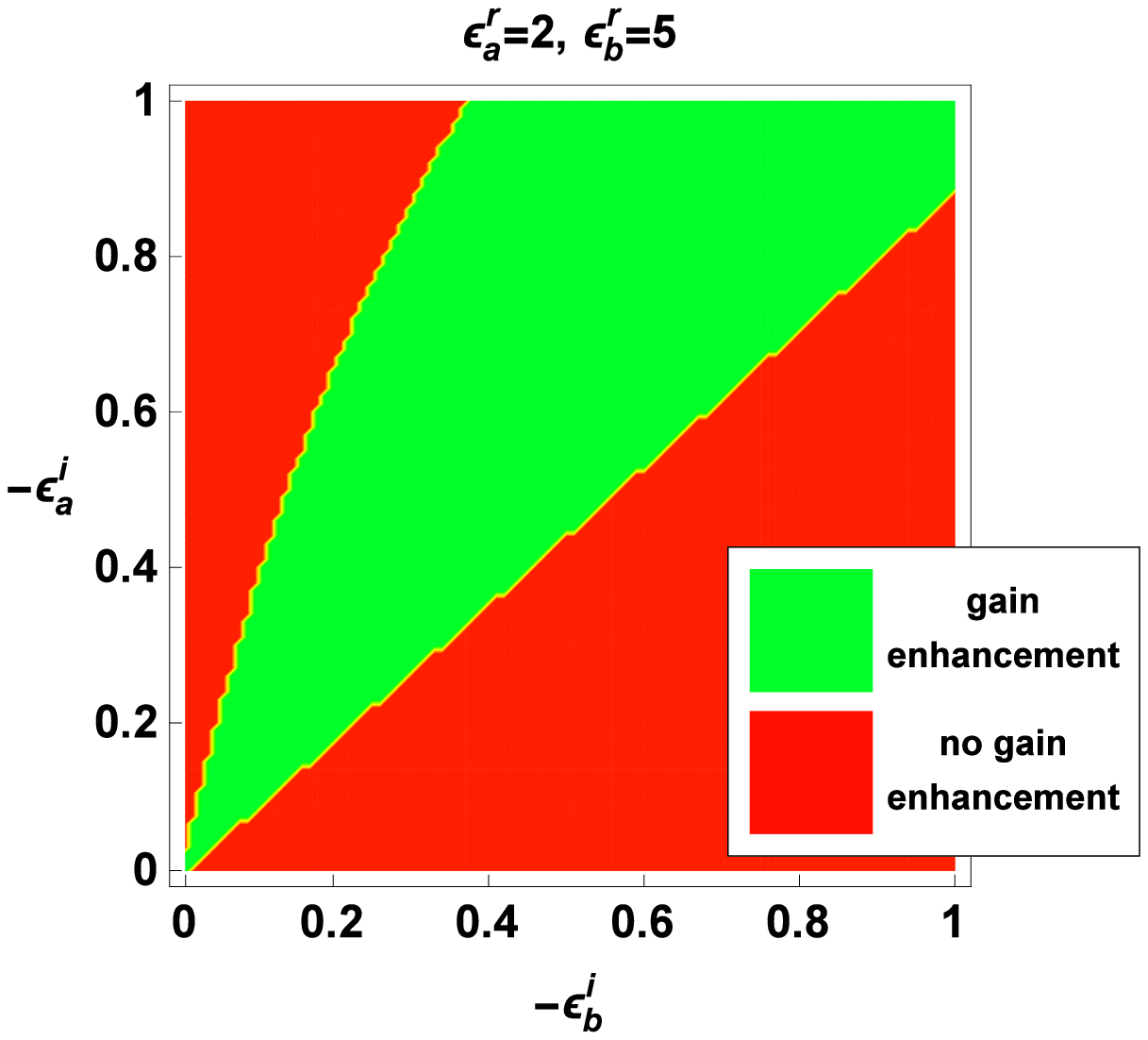}
\\
\includegraphics[width=12.0cm]{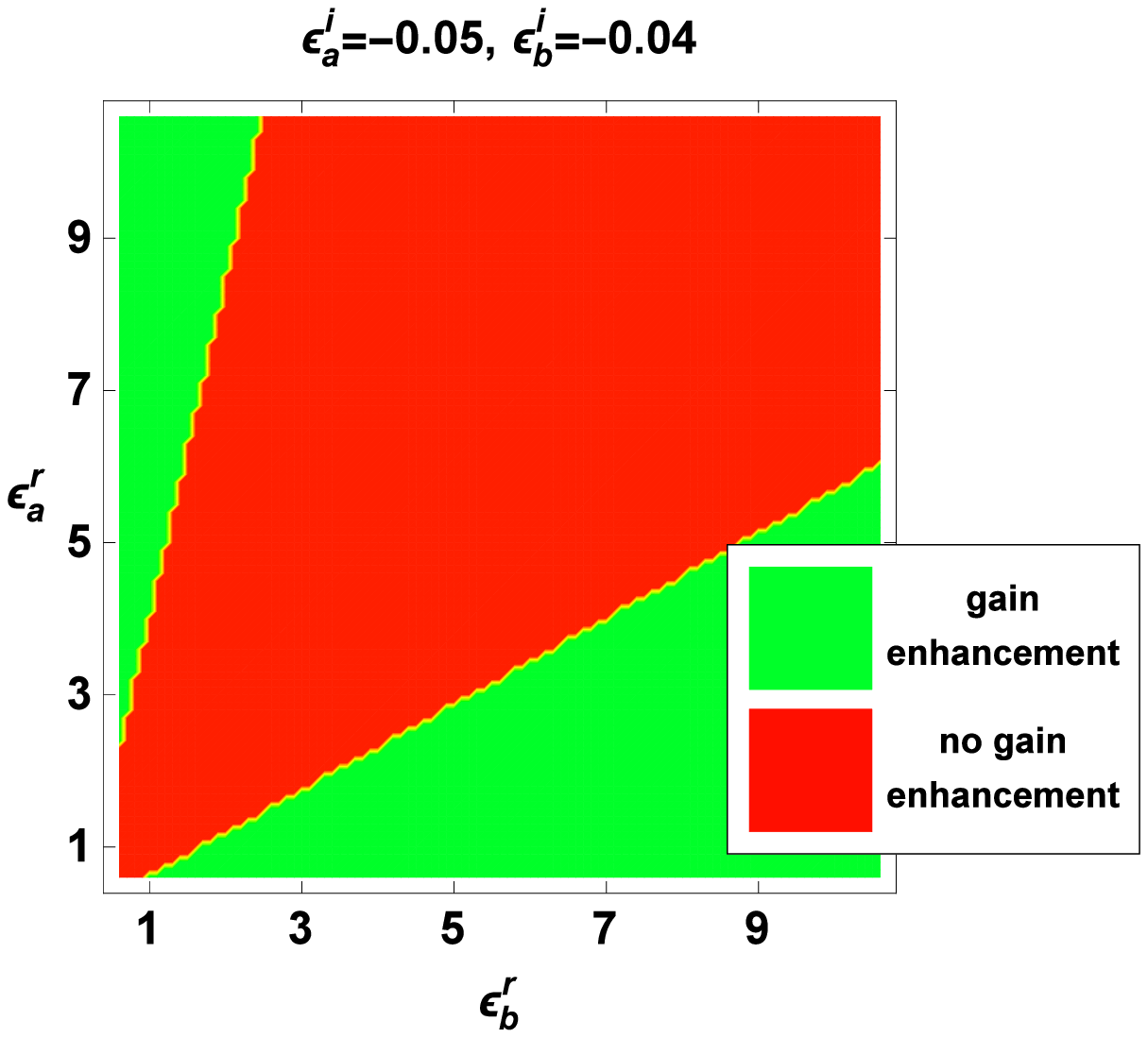}
\end{tabular}
\end{center}
 \caption{Top: Gain-enhancement subspace in the $\le -\eps^i_a, -\eps^i_b \ri$ space,
 when
  $\eps^r_a = 2$ and $\eps^r_b = 5$. Bottom: Gain-enhancement subspaces
in the $\le \eps^r_a, \eps^r_b \ri$ space, when
 with $\eps^i_a = - 0.05$ and $\eps^r_b = -0.04$. } \label{fig5}
\end{figure}

In Fig.~\ref{fig5}, the parameter spaces  that support gain enhancement are mapped for: (i) $\le -\eps^i_a, -\eps^i_b \ri \in \le 0, 1 \ri \times \le 0, 1 \ri $
 with $\eps^r_a = 2$ and $\eps^r_b = 5$; and (ii)
  $\le \eps^r_a, \eps^r_b \ri \in \le 0.5, 10 \ri \times \le 0.5, 10 \ri $
 with $\eps^i_a = - 0.05$ and $\eps^r_b = -0.04$.
 For  $\eps^r_a = 2$ and $\eps^r_b = 5$, the  gain-enhancement subspace in the $\le -\eps^i_a, -\eps^i_b \ri$ space is a  window that contains $\eps^i_a = \eps^i_b$ and   becomes narrower as the magnitudes of
 $\eps^i_a$ and $\eps^i_b$ are decreased. For
  $\eps^i_a = -0.05$ and $\eps^i_b = -0.04$,  two  gain-enhancement subspaces in the $\le \eps^r_a, \eps^R_b \ri$ space exist where $\eps^r_a$ and $\eps^r_b$ are dissimilar in magnitude with greater scope for gain enhancement arising when the magnitudes of $\eps^r_a$ and $\eps^r_b$ are increased.
These trends gleaned from Fig.~\ref{fig5} are wholly consistent with those evident in Figs.~\ref{fig3} and \ref{fig4}.

\subsection{Non-dissipative and non-active component materials}
In passing, let us remark on the special case when both component materials are neither
dissipative nor active, i.e., $\eps^i_a = \eps^i_b = 0$.
 Provided that the possibility  $\eps_{Br} = 0$ is excluded from  consideration (which is not physically plausible for the situation
  $\eps_a \eps_b >0 $ considered here),
 we  infer from Eq.~\r{dBr} that $ \displaystyle{ {d \eps_{Br}}/{d f_a} \neq 0}$.
Therefore, $\eps_{Br}$ is either a uniformly increasing or a uniformly decreasing function of $f_a$.  Hence,
$\eps_{Br}$ must lie between $\eps_a$ and $\eps_b$ for all values of $f_a\in[0,1]$.

\section{Closing remarks}

Using the Bruggeman formalism, we have established in the foregoing sections that
an HCM comprising
two active (resp. dissipative) component materials may exhibit more gain (resp. loss) than either of its component materials.
For the range of $\eps_a$ and $\eps_b$ values explored in numerical examples here, gain enhancements of up to 40\% were found.
Furthermore, sufficient conditions for such gain enhancement and loss enhancement have been established in conditions
 \r{cond1}  and \r{xcond1}, and  \r{cond4} and \r{cond5}, respectively.
These enhancements  arise when  (i) the imaginary parts of the relative permittivities of both component materials are similar in magnitude and (ii) the real parts of the relative permittivities of both component materials are dissimilar in magnitude.
 Similar gain/loss enhancements also
emerge from the Maxwell Garnett formalism for dilute composite materials.

The reported phenomenons of gain enhancement and loss enhancement are likely to be exacerbated by directional effects in anisotropic HCMs, as has been established for  nonlinearity enhancement \c{LL01,M03} and group-velocity enhancement \c{ML_group_vel_2}.

\vspace{12mm}

\noindent {\bf Acknowledgments:} TGM acknowledges the support of EPSRC grant EP/M018075/1.
AL thanks the Charles Godfrey
Binder Endowment at Penn State for ongoing financial support of his
research activities.

\newpage

\newpage

\newpage

\newpage

\newpage

\begin{thebibliography}{99}



\bibitem{L96}  Lakhtakia A (ed.).
Selected papers on linear optical composite materials. Bellingham (WA): SPIE
Optical Engineering Press; 1996.


\bibitem{Walser}
  Walser RM.   Metamaterials: an introduction. In:
Introduction to complex mediums for optics and
electromagnetics. (W.S. Weiglhofer and A. Lakhtakia, eds). Bellingham (WA): SPIE
Press;  2003, pp.~295--316. 

\bibitem{Meta}
Cui TJ, Smith D, and Liu R (eds.).
Metamaterials: theory, design, and applications.
New York (NY): Springer; 2010.

\bibitem{Boyd96}
  Boyd RW,   Gehr RJ, Fischer GL, and  Sipe JE.  Nonlinear optical properties of nanocomposite materials. Pure Appl. Opt.
  1996;5:505--512. 


\bibitem{Liao}
  Liao HB,   Xiao RF,  Wang H,   Wong KS, and  Wong GKL.
Large third--order optical nonlinearity in  $ \mbox{Au:TiO}_2$ composite films
measured on a femtosecond time scale. Appl. Phys. Lett. 1998;72:1817--1819. 


\bibitem{L01}
 Lakhtakia A.  Application of strong  permittivity fluctuation theory for isotropic, cubically nonlinear, composite mediums.
Opt. Commun. 2001;192:145--151.

\bibitem{Solna2}
  S{\o}lna K and   Milton GW. Can mixing materials make electromagnetic signals travel faster?  SIAM J. Appl. Math. 2002;62:2064--2091. 

\bibitem{ML_group_vel_1}
 Mackay TG and  Lakhtakia A.
 Enhanced group velocity in metamaterials.
J. Phys. A: Math.  Gen. 2004;37:L19--L24. 

\bibitem{Gao_06}
 Gao L. Decreased group velocity in compositionally graded films.
Phys.  Rev. E 2006;73:036602. 


 \bibitem{Br}
 Bruggeman  DAG.
 Berechnung verschiedener physikalischer Konstanten von heterogenen Substanzen, I.
Dielektrizit\"ats\-konstanten und Leit\-f\"ahig\-keiten der
Misch\-k\"orper aus isotropen Substanzen. Ann. Phys. Lpz. 1935;24:636--679. (Reproduced in \c{L96}).

\bibitem{Aspnes}
 Aspnes DE.
Local--field effects and effective--medium theory: a microscopic
perspective.
Am. J. Phys.  1982;50:704--709.  (Reproduced in \c{L96}). 

\bibitem{MAEH}
 Mackay TG and  Lakhtakia A.
 Modern  analytical electromagnetic homogenization.
Bristol (UK): IOP Publishing; 2015.



\bibitem{ML_Br_limitation}
 Mackay TG and  Lakhtakia A. A limitation
of the Bruggeman formalism for homogenization. Opt.
Commun. 2004;234:35--42. Erratum 2009;282:4028.


\bibitem{Wiener}
Wiener O.
Die Theorie des Mischk\"orpers f\"ur das Feld der
Station\"aren Str\"omung.
Abh. Math.--Phys. Kl. S\"achs.
 1912;32:507--604.  (Reproduced in \c{L96}, which also provides an English synopsis of \c{Wiener} by B. Michel).



\bibitem{HS}  Hashin Z and   Shtrikman S.
 A variational approach to the theory of the effective magnetic
permeability of multiphase materials. J. Appl. Phys. 1962;33:3125--3131. 


\bibitem{Milton}
Milton GW.
Bounds on the complex dielectric constant
of a composite material.
Appl. Phys. Lett. 1980;37:300--302. 

\bibitem{MG}
Maxwell Garnett JC.
Colours in metal glasses and in metallic films.
 Phil. Trans. R. Soc. Lond. A 1904;203:385--420. (Reproduced in \c{L96}).


\bibitem{LL01}
Lakhtakia MN and   Lakhtakia A.
Anisotropic composite materials with intensity--dependent permittivity tensor: the Bruggeman
approach. Electromagnetics 2001;21:129--138.

\bibitem{M03}
 Mackay TG. Geometrically derived anisotropy in cubically nonlinear dielectric composites. J. Phys. D: Appl. Phys. 2003;36:583--591.


\bibitem{ML_group_vel_2}
Mackay TG and  Lakhtakia A. Anisotropic enhancement of group velocity in a homogenized dielectric composite medium. J. Opt. A: Pure  Appl. Opt. 2005:7;669--674. 


\end{thebibliography}
\end{document}